\documentclass[mathleft]{an}
\usepackage{graphicx}
\usepackage{times}
\usepackage{natbib}
\newcommand{\reff}{r_\mathrm{eff}}

\newcommand{\zform}{z_\mathrm{form}}

\newcommand{\rhodm}{\langle \rho_\mathrm{DM} \rangle}
\newcommand{\der}{\mathrm{d}}
\newcommand{\piz}{\Pi_{zz}}

\newcommand{\pir}{\Pi_{RR}}
\newcommand{\pip}{\Pi_{\phi\phi}}

\begin{document}

\Pagespan{1}{}
\Yearpublication{2010}%
\Yearsubmission{2009}%
\Month{12}%
\Volume{999}%
\Issue{88}%

\title{Schwarzschild modelling of elliptical galaxies and their black holes}

\author{Jens Thomas\inst{1,2}\fnmsep\thanks{\email{jthomas@mpe.mpg.de}\newline}
}

\titlerunning{Dynamical modelling of ellipticals}
\authorrunning{J. Thomas}
\institute{Universit\"ats-Sternwarte M\"unchen, Scheinerstr. 1, 
D-81679 M\"unchen, Germany
\and 
Max-Planck-Institut f\"ur extraterrestrische Physik,
Giessenbachstr., 85748 Garching, Germany
}

\received{30 May 2005}
\accepted{11 Nov 2005}
\publonline{later}


\abstract{This article describes the Schwarzschild orbit superposition method. It is the
state-of-the-art dynamical modelling tool for early-type galaxies. Tests with analytic
models show that masses and orbital anisotropies of not too face-on galaxies 
can be recovered with about 15 percent accuracy from typical observational data. Applying 
Schwarzschild models 
to a sample of Coma galaxies their dark matter halos were found to be 13 times denser 
than those of spirals with the
same stellar mass. Since denser halos assembled earlier, this result indicates 
that the formation redshift $1+z_\mathrm{form}$ of ellipticals is about two times
higher than of spirals. Roughly half of the sample galaxies have halo assembly redshifts 
in agreement with their stellar-population ages. Galaxies where stars appear younger 
than the halos show strong phase-space density gradients in their orbital structure, 
indicative for dissipational evolution and possibly connected with secondary star-formation 
after the main halo assembly epoch. The importance of considering dark-matter in dynamical 
models aimed to measure black-hole masses is briefly discussed.}

\maketitle

\section{Introduction}
Early-type galaxies are characterised by an overall smooth and featureless
spheroidal morphology and a dynamically hot system of stellar orbits. This is
thought to be the result of a dynamically violent assembly process. However, when and
how exactly these galaxies have formed is still not well known. Their mostly
old and $\alpha$-enhanced stellar populations imply a relatively short star-formation 
period in the distant past. The epoch when those stars assembled to form the early-types 
seen today cannot be directly deduced from stellar population
ages. For example, the stars might be born at high redshift, in progenitors that only 
recently merged into spheroidal galaxies. If these mergers are mostly collisionless (without 
significant amounts of gas and star-formation), then stellar-population ages do not change,
and the main assembly epoch is delayed with respect to the star-formation epoch. In contrast, 
a spheroidal galaxy formed early in the universe could have grown
a disky stellar subcomponent recently, e.g. triggered by a gas-rich minor merger.
The main halo assembly would then precede star formation (along the disk).

Pure dark-matter, collisionless $N$-body simulations predict a close relationship between 
the main assembly redshift of dark matter halos (when, say, half of the mass had been 
assembled) and their average density. The infall of baryons might change the dark matter
density as it enforces an extra gravitational pull on the halo. Still,
measuring the dark matter density in elliptical galaxy halos is a valuable tool
to gain information about their assembly epoch. In comparison with stellar-population ages 
it also gives indirect information about the evolution of these systems. According to the
above considerations one can expect that galaxies in which stars appear younger than the
halos have likely experienced some secondary star-formation episodes. These systems have
evolved dissipatively. In contrast, when stars appear older than the halo, this might
indicate a more gasless evolution (e.g. by gas-free or collisionless or dry, respectively, 
mergers).

Since the system of stars in galaxies is collisionless it preserves some information
about how the stars have assembled. A dynamically
violent formation -- characterised by violent relaxation in phase-space --
is supposed to result in a highly mixed orbit distribution with strong phase-space
density gradients likely being washed out. In contrast, dissipational evolution (e.g. through
gas-rich mergers) likely results in disky subsystems with high 
phase-space density peaks on near-circular orbits. Consequently, 
the analysis of the orbital structure provides additional information about 
the assembly mechanism of early-type stellar systems.

In the last years we have collected photometric and kinematic observations for
a sample of 19 early-type galaxies in the Coma cluster: 2 central cD galaxies, ten giant 
ellipticals and seven S0 or E/S0 galaxies, respectively, with $-18.8 \ge M_B \ge -22.26$
(Mehlert et al. 2000; Wegner et al. 2002; Corsini et al. 2008; Thomas et al. 2009a). 
For all galaxies ground-based and HST photometry
are available. Long-slit stellar absorption-line spectra have been taken along at
least the major and minor axes. In many cases additional data along other position
angles has been collected as well. The spectra extend to $1-4\,\reff$, entering the regions
where dark matter becomes noticeable.

By means of dynamical models we have measured the dark matter content and orbital structure 
of these galaxies. The results are summarised below 
and implications for the formation process of early-type galaxies are discussed.

\section{The Schwarzschild method}
A complete
description of a stellar system is provided by its phase-space distribution function $f$, i.e.
the density of stars in 6-dim phase-space. Unlike for a collisional gas the distribution 
function of collisionless dynamical systems like 
galaxies is not known in advance. However, for steady-state
objects Jeans theorem ensures that the phase-space density is constant along individual
orbits. Orbits, in turn, are identified by the integrals of motion they respect. Thus,
for systems in a steady state the distribution function (DF) reads 
\begin{equation}
f = f(I_1,\ldots,I_n),
\end{equation}
 where $I_1,\ldots,I_n$ are
integrals of motion (e.g. Binney \& Tremaine 1987). The simplest symmetry assumption 
consistent with the flattening and rotation of elliptical galaxies is that they are
axisymmetric. Orbits in typical axisymmetric 
galaxy potentials respect three integrals of motion: energy $E$, angular momentum along 
the symmetry axis $L_z$ (the rotation axis being parallel to the $z$-axis) and 
the so-called third integral $I_3$ (e.g. Contopoulos 1963). The last integral is 
usually not known explicitly and
the distribution function can thus not be written in terms of elementary functions. 

Schwarzschild (1979) introduced an
orbit-superposition technique (now called Schwarzschild method) to construct
collisionless DFs. In brief,
one assumes a trial potential for a given galaxy and composes a library of several
thousand orbits. The light distribution and projected kinematics of each orbit are
stored. Then, a galaxy model is constructed as the superposition of all orbits. The
unknown weight or total amount of light, respectively, of each orbit is chosen 
to match the model as good as possible with the given constraints (see below). This method
corresponds to the approximation $f \approx \sum_i f_i$, where the $f_i$ are single-orbit 
DFs (Vandervoort 1984; Thomas et al. 2004). The accuracy of the method only depends 
on the density of the orbit grid. 

The Schwarzschild method provides very general dynamical models. In contrast
to Jeans models there is no necessity for any a priori restriction upon 
the orbital structure.
Moreover, Schwarzschild models are easily constructed to be everywhere
positive in phase space (i.e. physically meaningful) -- a property that is not guaranteed
in Jeans models. 
The main challenge is to ensure that the orbit library is representative 
for all orbital shapes. 

In axisymmetric potentials it is straight forward to 
sample the energy $E$ and the angular momentum $L_z$. 
The issue related to our ignorance about 
$I_3$ is overcome by a systematic sampling of orbital initial conditions, which implicitly 
guarantees the inclusion of all orbital shapes. More specifically,
orbits with given $E$ and $L_z$ but different $I_3$ follow 
distinct invariant curves in 
appropriate surfaces of section (SOS). Launching orbits at constant $E$ and $L_z$ from 
various initial starting points on a grid in such SOSs ensures
a representative sampling of the unknown $I_3$ (Thomas et al. 2004).

In practice, one goes through the following steps to construct a Schwarzschild model
of a galaxy:
\begin{itemize}
\item
The photometry of the galaxy is deprojected to obtain the 3d light distribution $\nu$.
\item
A trial mass distribution $\rho$ is set up, combining the various mass components, e.g.
\begin{equation}
\rho = \Upsilon \times \nu + \rho_\mathrm{DM} + M_\mathrm{BH} \times \delta(r).
\end{equation}
Here, $\Upsilon$ is the stellar mass-to-light ratio (typically assumed to be radially 
constant in ellipticals), $\rho_\mathrm{DM}$ is the dark matter density (see below) and
$M_\mathrm{BH}$ represents a supermassive central black hole. Given $\rho$, the
gravitational potential follows by solving Poisson's equation and an orbit
library can be assembled.
\item
The orbits are superposed and the orbital weights that best match
with the observations are determined. In our implementation this is done by
maximising 
\begin{equation}
\label{orbitmaxs}
S - \alpha \chi^2,
\end{equation}
where $\chi^2$ quantifies deviations between observed and modelled kinematics (the
observed light profile is used as a boundary condition in the maximisation). The
function
\begin{equation}
\label{maxs}
S = - \int f \ln f \, \der^3 r \, \der^3 v 
\end{equation}
is the Boltzmann entropy of the orbit distribution and $\alpha$ is a regularisation
parameter (Richstone \& Tremaine 1988; see also below).

\item
The parameters determining the potential (e.g. $\Upsilon$, halo parameters,
$M_\mathrm{BH}$) are systematically varied and the final best-fit is obtained from a 
$\chi^2$ analysis.
\end{itemize}
Various implementations of this method exist for spherical (Romanowsky et al. 2003), 
axisymmetric 
(Cretton et al. 1999; Cappellari et al. 2006; Chanam{\'e}, Kleyna \& Van der Marel 2008; 
Valluri, Merritt \& Emsellem 2004; Gebhardt et al. 2003; Thomas et al. 2004) as-well as 
triaxial potentials (van den Bosch et al. 2008). Gebhardt et al. (2003) measured 
supermassive black-holes in the centres of galaxies 
(ignoring any contribution of dark matter). Cappellari et al. (2006) modelled a subsample of 
the SAURON galaxies, again ignoring dark matter because their data cover only the inner 
regions $<\reff$.

Beyond $\reff$ dark matter becomes important in early-type galaxies. The Coma sample
is by now the only larger sample of generic (e.g. flattened
and rotating) ellipticals that has been modelled with Schwarzschild's method including
dark matter. Previous attempts to measure the dark matter content of early-types via
stellar dynamics focussed on round and non-rotating systems and assumed spherical symmetry
(Kronawitter et al. 2000; Gerhard et al. 2001; Magorrian \& Ballantyne 2001). 

In the Coma galaxies we probed for two
different parametric halo profiles: (1) logarithmic halos with a constant-density 
core and asymptotically constant circular velocity and (2) so-called NFW-profiles 
which are good fits to dark matter halos in cosmological $N$-body 
simulations (Navarro, Frenk \& White 1996). The majority of Coma galaxies are better fit 
with logarithmic halos, but the significance over NFW halo profiles is marginal (Thomas et al.
2007b).

\section{Tests of the method}
\begin{figure}
\includegraphics[width=82mm]{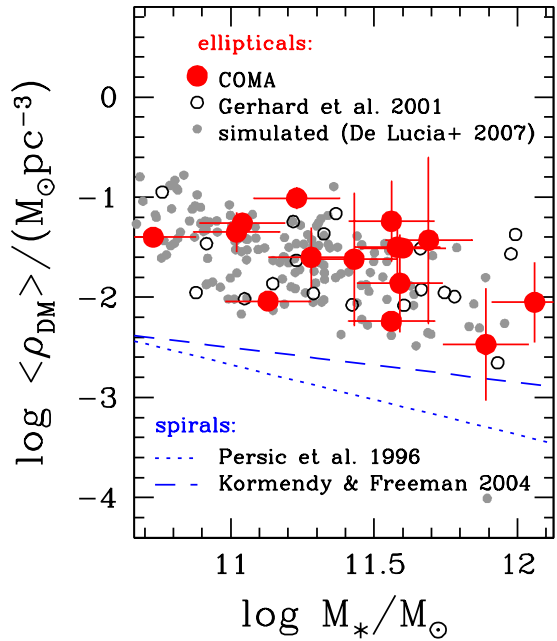}
\caption{Dark matter density $\rhodm$ (averaged within $2 \, \reff$) versus stellar mass
$M_\ast$. Red dots:
Coma early-type galaxies; open circles: round and non-rotating galaxies from Gerhard et al
(2001); small dots: semi-analytic galaxy formation models (De Lucia \& Blaizot 2007).
Blue lines: spiral galaxies.}
\label{fig:rhodm}
\end{figure}
A unique feature of our Schwarzschild models is the incorporation
of orbital phase-space volumes $V$ (Thomas et al. 2004). Together with the total 
amount of light $w$ on the orbit it allows to calculate the phase-space density
$f= w/V$ on each orbit. This offers several new applications. Firstly, one
can project analytic phase-space DFs $f_\mathrm{analytic}$ through orbit 
libraries. In this case,
the orbital weights are not determined from a fit to photometric and kinematic constraints,
but directly from the analytic DF, i.e. the weight $w_i$ of orbit $i$ reads 
$w_i = f_\mathrm{analytic} \times V_i$. The projected kinematics,
spatial density profile, and intrinsic velocity dispersion profiles of the so-constructed
orbit superposition can be compared with direct phase-space integrations of 
$f_\mathrm{analytic}$. Both methods agree very well (Thomas et al. 2004), confirming that 
the orbit sampling is representative.

In addition, mock data from analytic model galaxies can be used to measure how
accurate the orbital DF can be reconstructed. In this Monte-Carlo approach photometric and 
kinematic data with spatial resolution and coverage similar to real data are 
simulated and modelled exactly as real galaxies. For the Coma data we have shown 
that with the appropriate choice of the regularisation parameter $\alpha$ the orbital
structure and mass distribution of not too face-on galaxies 
can be recovered with an accuracy of about 15 percent
(Thomas et al. 2005). Similar tests have been presented in Cretton et al. (1999), 
Krajnovi\'c et al. (2005) and Siopis et al. (2009).

The input models for the above tests had the same symmetry as the dynamical
models (axial symmetry). Isophotal twists and kinematic misalignment
suggest that at least the most massive early-type galaxies are slightly triaxial.
In order to explore the systematic errors arising from too restrictive symmetry 
assumptions, we applied our models also to mock data sets created from 
collisionless $N$-body binary disk merger remnants. These remnants are strongly
triaxial in the central, box-orbit dominated regions (Jesseit, Naab \& Burkert 2005). 
While the central mass measured with axisymmetric
models then underestimates the true mass on average (depending on projection angle and 
box-orbit content) the enclosed mass within $\reff$
is still recovered mostly with better than 20 percent accuracy unless for highly
flattened, face-on systems (Thomas et al. 2007a).

\section{The dark matter density and assembly epoch of early-type galaxies}
Fig.~\ref{fig:rhodm} shows dark matter densities $\rhodm$ as a function of stellar 
mass. Stellar masses $M_\ast = \Upsilon \times L$ of early-types are derived 
from the best-fit (dynamical) stellar mass-to-light ratio $\Upsilon$ and the observed total 
luminosity $L$. Dark matter densities $\rhodm$ are averaged within $2 \, \reff$.

The dark matter densities of Coma ellipticals (from axisymmetric  modelling) and round and 
non-rotating galaxies (from spherical models) match well. With increasing stellar mass
dark matter densities tend to decrease. Spiral galaxy dark matter densities are
also shown in Fig.~\ref{fig:rhodm} (Persic, Salucci \& Stel 1996a,b; Kormendy \&
Freeman 2004, respectively). 
Stellar masses for spirals have been derived using the Tully-Fisher and stellar-mass 
Tully-Fisher relations of Bell \& De Jong (2001) (cf. Thomas et al. 2009a for details). As in
early-types, dark matter densities in spirals decrease with increasing stellar mass. 
Most important, the dark matter in ellipticals is about $13$ 
times denser than in spirals of the same stellar mass (compared at the same $B$-band 
luminosity, early-type galaxies have $7$ times denser halos than late-types).

Finally, Fig.~\ref{fig:rhodm} also includes semi-analytic galaxy formation 
models of De Lucia \& Blaizot (2007). They are in good
agreement with the observations in Coma galaxies. This is surprising because
the $N$-body cosmological simulation underlying the semi-analytic models 
was performed without baryon dynamics. Thus,
either the net effect of baryons on dark matter around early-types in the given
mass interval vanishes or there is some discrepancy between the semi-analytic models and the
observations. Note that baryons can also cause a halo expansion (e.g.
by dynamical friction).

\begin{figure}
\includegraphics[width=82mm]{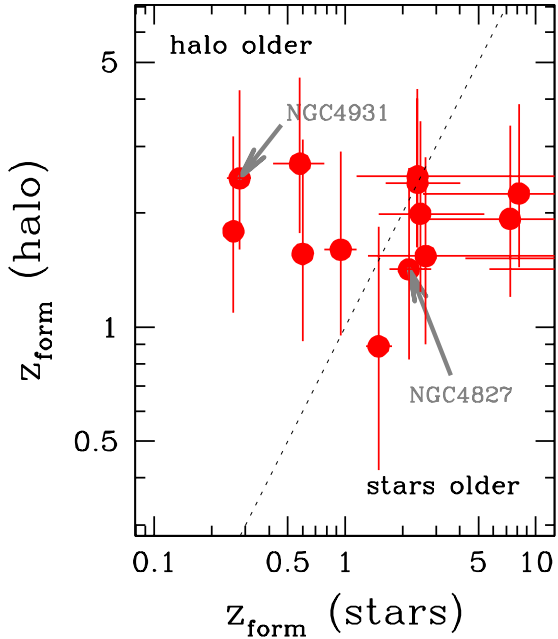}
\caption{Halo assembly redshifts (y-axis) versus star-formation redshift
(x-axis) for Coma early-type galaxies. The one-to-one relation is indicated
by the dotted line. The phase-space DFs of the two
galaxies NGC4827 and NGC4931 are shown in Fig.~\ref{fig:df}.}
\label{fig:zz}
\end{figure}

The light distribution in late-type galaxies is less centrally concentrated
than in early-types. Then, even if the net effect of baryons on elliptical galaxy 
halos would be negligible, the stronger gravitational pull in early-type galaxy
halos could still contribute to the observed 
overdensity of dark matter in ellipticals relative to spirals. Within the adiabatic contraction
approximation this can at most explain a 
factor of two between average halo densities around ellipticals and spirals, respectively 
(Thomas et al. 2009a). The remaining over-density indicates
that ellipticals have assembled earlier -- at a time where the universe was denser. 

The simplest analytic models, as well as pure dark matter cosmological $N$-body
simulations predict a scaling of the average dark matter density $\rhodm$ 
with halo-assembly redshift $\zform$ of $\rhodm \propto (1+\zform)^3$
(e.g. Gerhard et al. 2001). Accordingly, the density contrast
between elliptical and spiral galaxy halos translates into $1+\zform$ being about two 
times higher for ellipticals than for spirals. Absolute assembly redshifts can only 
be estimated with an additional assumption upon the formation redshift of spirals. 
Supposed that spirals assemble typically at $\zform \approx 1$
(at higher redshifts spirals become rare, e.g. Conselice et al. 2005) then
Coma early-types have assembled around $\zform \approx 2-3$.
These dark-matter based formation redshifts are shown in 
Fig.~\ref{fig:zz} against stellar-population ages. In about one third of Coma galaxies 
the stars appear to be younger than the
halos but many galaxies are consistent with equal assembly and star-formation redshifts.

\section{The orbital structure of early-type galaxies}
As outlined above a galaxy in which the stars appear to be younger than the halo is a candidate dissipative
system, while objects in which stars are equally old or older than the halo have likely
formed monolithically or through mostly collisionless mergers. In dissipative systems one
would expect a high phase-space density on near-circular disk orbits while the phase-space
DF of systems which have undergone violent relaxation should lack strong phase-space
density gradients.

Classically,
the orbital structure is measured in terms of anisotropy parameters, i.e. ratios of 
intrinsic velocity dispersions. Let $f$ denote the phase-space DF of a
galaxy, then the intrinsic dispersions $\sigma_{ij}$ read
\begin{equation}
\label{e:sig}
\sigma_{ij}^2 = \frac{1}{\rho} \, \int f \, (v_i-\overline{v_i})(v_j-\overline{v_j}) \, \der^3 v,
\end{equation}
\begin{equation}
\label{e:vav}
\overline{v_i} = \frac{1}{\rho} \, \int f \, v_i \, \der^3 v.
\end{equation}
In the following we assume $i,j \in \{R,\phi,z\}$, 
where $z$ is the symmetry axis (cylindrical coordinates).
The unordered kinetic energy along coordinate direction $i$ is 
\begin{equation}
\label{pidef}
\Pi_{ii}=\int \rho \sigma_{ii}^2 \, \der^3 r
\end{equation}
and the anisotropy in the velocity dispersions can be quantified by
$\beta \equiv 1 - \piz/\pir$ and
$\gamma \equiv 1 - \pip/\pir$ (Cappellari et al. 2007). An isotropic system
has $\beta = \gamma = 0$.

\begin{figure}
\includegraphics[width=82mm]{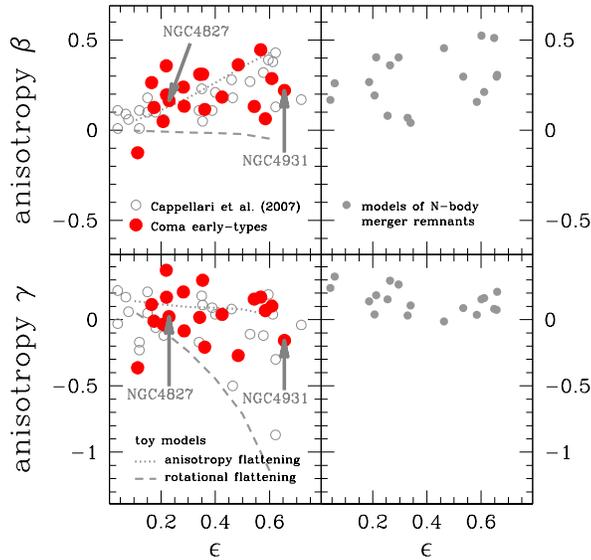}
\caption{Anisotropy $\beta$ and $\gamma$ versus intrinsic ellipticity $\epsilon$ 
for observed galaxies
(left-hand panels) and $N$-body binary disk merger remnants (right-hand panels). Dotted
lines: maximum-entropy toy-models, dashed lines: 2I models (details in the text). The
phase-space DFs of the two galaxies NGC4827 and NGC4931 are shown in Fig.~\ref{fig:df}.}
\label{fig:aniso}
\end{figure}

Fig.~\ref{fig:aniso} shows $\beta$ and $\gamma$ versus intrinsic ellipticity $\epsilon$.
In observed galaxies $\beta > 0$ and, on average, $\gamma \approx 0$, but with significant
scatter. The range of anisotropies found in Coma early-types is similar as in SAURON 
galaxies but the Coma galaxies do not exhibit a trend of increasing $\beta$ with
increasing $\epsilon$ as observed by Cappellari et al (2007). Neither the inclusion of dark matter
in the Coma models nor regularisation can explain this difference (Thomas et al. 2009b). Instead it 
most likely reflects differences in the selections of the two samples (Thomas et al. 2009b).

In order to clarify the relationship between the classical anisotropy
and the intrinsic orbital structure two proto-typical orbital compositions
are shown by the lines in Fig.~\ref{fig:aniso}: (1) The case of flattening
by rotation, i.e. by extra-light on near-circular orbits. Circular orbits
mostly contribute to the kinetic energy in $\phi$ direction and $\gamma$ becomes negative. The more,
the flatter the galaxy is (dashed lines). (2) The dotted lines show toy models with maximum 
entropy\footnote{For both toy models 75 percent of the light 
was distributed on prograde orbits; cf. Thomas et al. (2009b) for details.}. Without any other conditions the maximisation of
Boltzmann's entropy yields a flat DF $f = \mathrm{const}$. This is altered as soon as the
luminosity distribution is used as a boundary condition. Still, maximising the entropy
yields, in a sense, the smoothest phase-space DF compatible with a given
density profile. It turns out that the classical notion of flattening by anisotropy (i.e. 
a suppression of vertical versus horizontal energy, with isotropy between $R$ and $\phi$) 
is closely related to the maximisation of entropy. Accordingly, systems along the 
maximum-entropy line, while having varying anisotropy $\beta$, are similar in that their 
phase-space distribution functions are smooth. 

\begin{figure}
\includegraphics[width=82mm]{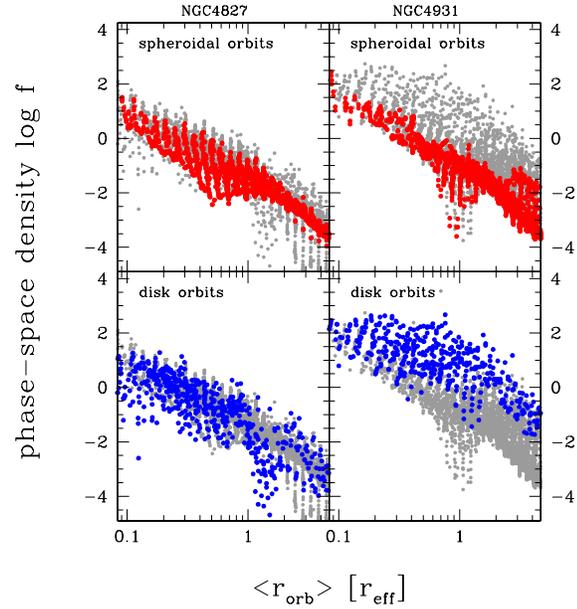}
\caption{Phase-space distribution function $f$ for the two galaxies NGC4827 (left-hand panels)
and NGC4931 (right-hand panels). Each dot represents the phase-space density on a single orbit 
(in solar masses per cubed kpc and cubed km/s) plotted against the mean orbital radius. 
In the top panels spheroidal orbits are
highlighted; in the bottom panel disk orbits are highlighted (details in the text). Only
prograde orbits are shown.}
\label{fig:df}
\end{figure}

The phase-space DFs of two Coma galaxies shown in Fig.~\ref{fig:df} (both 
galaxies are flagged by the arrows in Figs.~\ref{fig:zz} and \ref{fig:aniso}). The two galaxies have
similar $\beta$ and $\gamma$ but different intrinsic flattening: NGC4827 is an E2 early-type
while NGC4931 is highly flattened. In addition, in NGC4827 stars are about as old as the halo
while the stars in NGC4931 appear younger than its halo.

In the upper panels of Fig.~\ref{fig:df} spheroidal orbits are
highlighted. Disk orbits are highlighted in the lower panels.
We identify spheroidal orbits by $|\vartheta|_\mathrm{max} > 70\degr$, where
$\vartheta$ is the angle between a point on the orbit and the equatorial plane of the
galaxy. So-defined spheroidal orbits come close to the galaxy's minor-axis and contrast
disk orbits, which are instead selected by $|z|_\mathrm{max} < \reff/4$, where $z$ is the 
vertical height of a point on the orbit with respect to the equatorial plane. 

Fig.~\ref{fig:df} shows that phase-space densities on spheroidal orbits are similar
in both galaxies. Moreover, in NGC4827 the phase-space densities of disk orbits are
comparable to those of spheroidal orbits. Yet, the stellar density on disk orbits in NGC4931 
is up
to two orders of magnitude higher than on spheroidal orbits. These 
findings support the interpretation that galaxies left to the one-to-one relation in 
Fig.~\ref{fig:zz} (stars younger than halo) are dissipative systems, while galaxies close 
to the one-to-one relation or right of it
are systems which formed in a dynamically violent process and have smoother phase-space
DFs. A more detailed investigation of phase-space densities will be presented in 
Thomas et al. (in preparation).

The right-hand panels of Fig.~\ref{fig:aniso} display dynamical models of
collisionless $N$-body binary disk merger remnants. Mock data sets with
similar spatial resolution and coverage as for Coma galaxies were constructed from
projections along the three principal axes of six remnants selected representatively 
from the $N$-body sample of Naab \& Burkert (2003). While the distribution 
of $\beta$ versus $\epsilon$ is like in observed galaxies, none of the analysed 
merger remnants has $\gamma <0$. In view of the above discussion this is plausible since 
galaxies with $\gamma < 0$ are likely dissipative systems (large energy in $\phi$ caused
by extra-light on circular orbits), while the analysed $N$-body merger simulations 
are gas-free.

\section{Summary}
The Schwarzschild technique is the state-of-the-art tool to model early-type galaxies.
For typical observational data masses and orbital anisotropies can
be recovered with about 15 percent accuracy (for not too face-on systems). We have applied
Schwarzschild models to 19 Coma early-type
galaxies to measure the dark matter content and distribution of 
stellar orbits. By today, it is the largest sample of generic early-type galaxies with dynamical
models including dark matter. Dark matter densities in early-types are larger than in 
spirals of the same luminosity or mass. Extra gravitational pulling by the more concentrated
baryons in ellipticals is not sufficient to explain this over-density. Instead, the formation
redshift $1+z_\mathrm{form}$ of early-types is about two times higher than of spirals. Under
the assumption $z_\mathrm{form} \approx 1$ for spirals the Coma
early-types have formed around $z_\mathrm{form} \approx 2-3$. Observed dark matter densities
are in good agreement with recent semi-analytic galaxy formation models. In about half of the
sample galaxies halo assembly redshifts match with stellar population ages. In galaxies
where stars appear younger than the halo, the orbit distribution indicates 
dissipational evolution (i.e. strong phase-space density peaks on near-circular disk
orbits). This suggest that these galaxies had some secondary star-formation after the
main halo assembly epoch. 

When modelling the very central regions of early-types to study their supermassive central
black-holes it is important to include dark matter in the models, even if the central
parts itself are not dominated by dark matter. Neglect of the halo can result in an
overestimation of the stellar mass-to-light ratio which subsequently leads to an overestimation
of the central {\it stellar} mass. The latter is degenerate with the black-hole mass and neglect
of dark matter might then yield a too small black-hole. This has been illustrated for
M87, where the black-hole mass more than doubled after including dark matter in the 
models (Gebhardt \& Thomas 2009).
The effect is supposed to be strongest for the most massive galaxies since their light
profiles are shallow. Further galaxies have to be modelled in order to establish if 
the inclusion of dark matter in black-hole models can reduce the discrepancy between
the $\approx 10^9 \, M_\odot$ solar mass black-holes in the most massive nearby galaxies
and the $\approx 10^{10} \, M_\odot$ solar mass black-holes in high-redshift quasars.



\end{document}